\documentstyle[11pt,paspconf,epsf]{article}

\begin{document}

\title{Atmospheres of Giant Planets from Neptune to Gliese 229B}

\author{Mark S. Marley}
\affil{New Mexico State University; Department of Astronomy; 
PO Box 30001/Dept. 4500; Las Cruces NM 88003 USA}

% Notice that some of these authors have alternate affiliations, which
% are identified by the \altaffilmark after each name.  The actual alternate
% affiliation information is typeset in footnotes at the bottom of the
% first page, and the text itself is specified in \altaffiltext commands.
% There is a separate \altaffiltext for each alternate affiliation
% indicated above.

% The abstract is entered in a LaTeX "environment", designated with paired
% \begin{abstract} -- \end{abstract} commands.  Other environments are
% identified by the name in the curly braces.

% Poster authors ONLY may omit the abstract in order to gain a little
% more page space for the text of the poster.

\begin{abstract}  Stratospheric heating, condensation,
convective transport of non-equilibrium species, and deep
radiative energy transport are important processes in the atmospheres
of the solar jovian planets. They likely affect the atmospheres
of extrasolar giant planets and brown dwarfs as well.   Stratospheric
temperatures control thermal fluxes in strong molecular
bands, and may dramatically affect observed spectra.  Condensation
processes affect the appearance of an object, alter
the abundances of atmospheric species, and influence the fluxes
of both reflected and emitted radiation.  Convection can dredge up
non-thermochemical equilibrium species from the deep atmosphere
to the observable atmosphere.  Finally deep radiative zones
can limit the effectiveness of deep convection and alter the
boundary condition at which the planet radiates energy to space.
Here I discuss the role these processes may play in the atmospheres
of brown dwarfs and extrasolar giant planets.
\end{abstract}

% Keywords should be included, but they are not printed in the hardcopy.


% That's it for the front matter.  On to the main body of the paper.
% We'll only put in tutorial remarks at the beginning of each section
% so you can see entire sections together.

\section{Introduction}

Two decades of exploration of the outer solar system
has given us a sophisticated
perspective on the physical processes which shape the observable 
atmospheres of the jovian planets.  Indeed this exploration has taught
us a number of lessons about planetary atmospheres which we
can take with us as we leave the solar system and begin to study the
atmospheres of the extrasolar planets and brown dwarfs.
Here I discuss a selection of four processes which  play substantial roles in these atmospheres and have 
the potential to play important roles in the atmospheres of extrasolar giant planets and
brown dwarfs as well.  
These processes are stratospheric heating, condensation, convective transport
of non-thermochemical equilibrium species, and deep radiative energy transport.
I will discuss both lessons learned about these processes from
the atmospheres of the solar jovian planets as well as applications
and model results for extrasolar planets and brown dwarfs.

For clarity, I will refer to Jupiter, Saturn, Uranus, and
Neptune as the solar jovian or giant planets.  I will assume that the
objects detected in orbit around other stars by the Doppler
method are indeed Jupiter-mass jovian, as opposed to terrestrial,
planets and will refer to these objects as extrasolar giant
planets (EGPs). This is done simply for economy of expression and is not meant 
to have any bearing on the question of whether or not some of these objects
are in fact brown dwarfs.  The most massive and warmest object
considered here is Gliese 229B.

\section {Atmosphere Models}

Model results presented below  are obtained using the modeling procedures
of Marley et al. (1996, 1997).  Radiative-convective equilibrium temperature
profiles are computed for assumed gravities, internal heat fluxes, and incident
radiative fluxes.  The deposition of incident radiation as a function of vertical position in
the atmosphere is self-consistently computed assuming
the mean orbital radius of the planet and the spectral type of the primary. 
No incident flux is assumed for Gliese 229 B.  Temperature profiles for
all objects are obtained by assuming cloud-free atmospheres.

Molecular opacities are treated using the exponential  sum (or k-coefficient) method  (Goody et al. 1989; Lacis \& Oinas 1991).
This technique is widely used in the analysis of planetary  atmospheres and is
accurate (e.g. Grossman \& Grant 1994).  It should not be confused with 
the Opacity Distribution
Function (ODF) method.   The limiting factor in the model accuracy is knowledge
of the molecular opacities, particularly those of methane,
at high temperatures and pressures, not the radiative transfer technique.

Computed temperature profiles for a variety of objects are shown in Figure 1.  For
\begin{figure}
\plotone{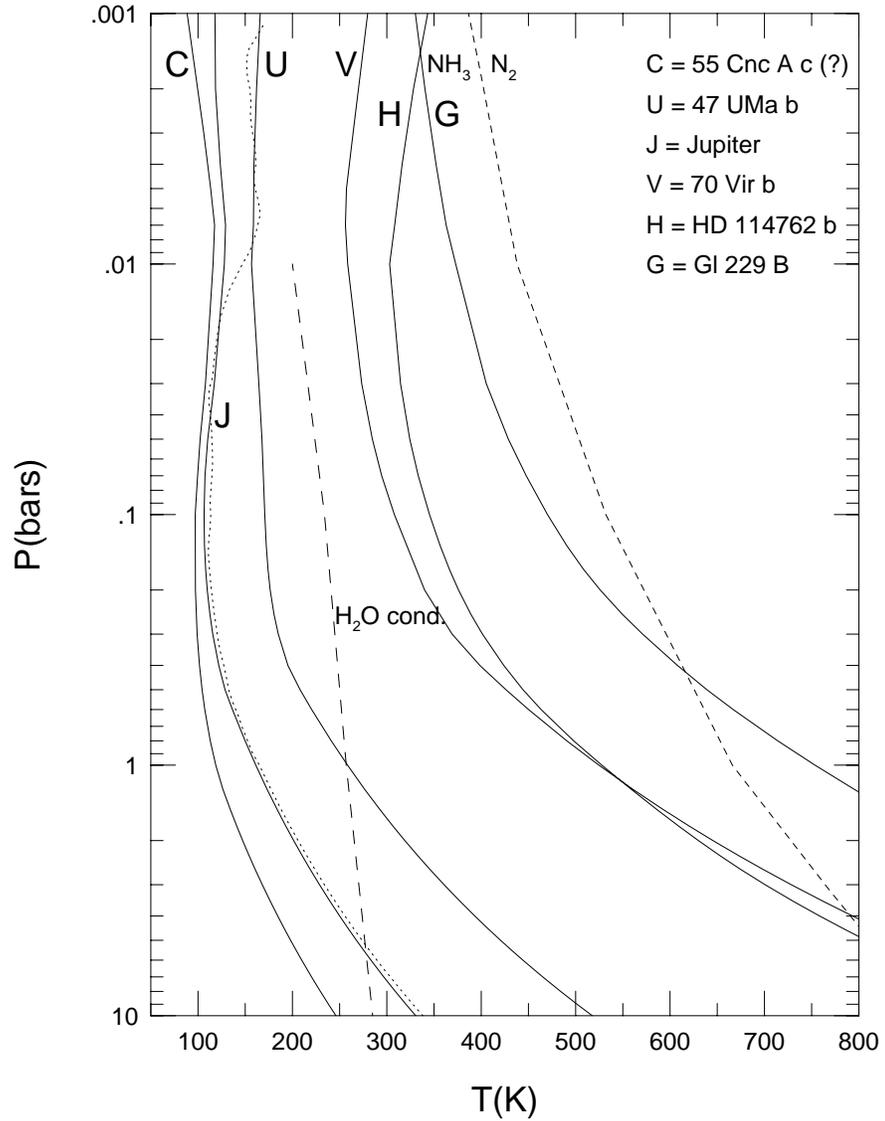}
\caption{Model temperature profiles for a variety of objects.  The
extrasolar planet 55 Cnc A c
is only suspected from residuals in the orbit of 55 Cnc A b (Butler et al.
1996).
Model Jupiter profile is computed in the same
way as the other profiles and is compared with observed profile (dotted).
Thermochemical equilibrium between N$_2$ and NH$_3$ is denoted by the short
dashed line; water condenses along the long dashed line.}
\label{fig-1}
\end{figure}
the EGPs, a typical model with parameters selected from a range of possible values for the surface
gravity and internal heat flux for the particular planet (Guillot et al. 1996) 
is shown.  The
accuracy of the approach is demonstrated by the close correspondence of the model and
actual temperature profiles for Jupiter (Lindal et al. 1981).  The Jupiter model
profile assumes solar composition
and no clouds.  Departure of the model from the observed profile  
at low pressure in the stratosphere is discussed in the next section.

\section{Stratospheric Heating}

The stratosphere is the region of a planetary atmosphere which is
in radiative equilibrium and lies immediately above the troposphere.
Unlike the temperature gradient in the troposphere which
falls with rising altitude, the stratospheric temperature is either constant
or increases with altitude.  A hot stratosphere overlying
a cold temperature minimum at the top of the troposphere (the tropopause)
is indeed a general feature of the solar planetary atmospheres.

If there were no incident radiation or other energy transport
mechanism, the stratosphere would have a temperature close to
the planetary ``skin'' temperature (e.g. Chamberlain and Hunten 1987).  
This is defined as the temperature
at zero optical depth in the thermal infrared,
$T_0 = 2^{-1/4} T_e$ where $T_e$ is the effective temperature.
In fact the stratospheric temperatures of the solar jovian planets,
as well as the Earth, all exceed the skin temperature. Thus the transport
of thermal radiation emitted by the lower atmosphere does not solely determine
the stratospheric temperature.

The discrepancy arises from the absorption of 
incident solar radiation within the stratosphere. On Earth the UV 
bands of ozone are 
responsible for stratospheric heating.  In the atmospheres of the
jovian planets the near-infrared methane bands play a similar role.
Hazes created by the condensation of photochemical products derived from
the decomposition of methane by solar UV radiation are also important. 
The influence of these hazes is complex 
as they  both scatter incident radiation out of the atmosphere
and absorb photons.  Other mechanisms, including wave breaking, may
also deposit energy and be responsible for some stratospheric heating.

The Jupiter model temperature profile shown in Figure 1 is derived
for a clear atmosphere.  The model does not include the radiative effect
of stratospheric hazes and is too cool compared to
the observed temperature profile in the stratosphere.  The warming
effect of hazes is detectable even at Neptune at 30 AU (Appleby 1986).  

The model  EGP profiles (Fig. 1) also show signs
of some stratospheric warming.  As with Jupiter, the models
likely underestimate the magnitude of this warming as no stratospheric
hazes are included.  In fact it is exceptionally difficult to model such hazes
on an a priori basis as their size and composition (and thus
radiative properties) depend upon a complex interplay of
eddy diffusion, condensation, and coagulation, which are 
peculiar to each individual object.  Even models
constrained to fit spacecraft observations have a difficult time
reproducing observed haze properties (e.g. Rages et al. 1991).

Gliese 229 A is sufficiently
dim and the brown dwarf lies at such large orbital distance that any stratospheric heating from
incident radiation would only amount to a few degrees and would likely be undetectable.   
Isolated brown dwarfs would also not be expected to exhibit stratospheric warming
unless there is a non-radiative energy transport mechanism (such as
wave breaking).

Even relatively modest stratospheric heating can produce large effects in the
thermal emission spectra.  
In particular the $7.8\,\rm \mu m$ methane band is a particularly 
effective stratospheric temperature
indicator.  In Jupiter's atmosphere emission from this strong bands originates in the
stratosphere.  The cold model stratosphere thus produces substantially less
flux in this band (Figure 2).  The flux at
\begin{figure}
\plotone{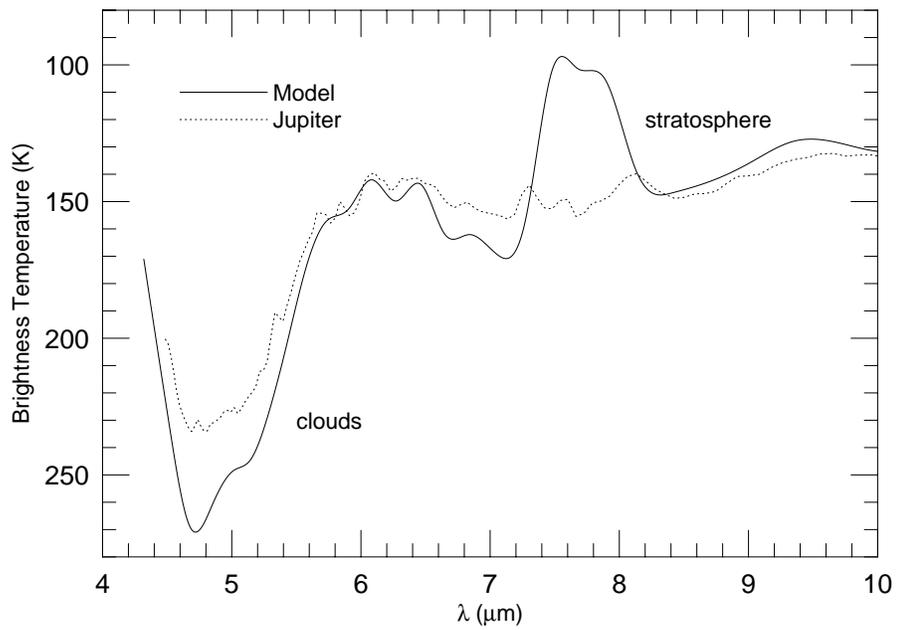}
\caption{Model and observed brightness temperature spectra for 
Jupiter.  
The cold model stratosphere (Figure 1) results in substantially
less flux in the $7.8\,\rm \mu m$ methane band than is the case
in the real Jupiter.  The lack of cloud opacity in 
the $5\,\rm \mu m$ spectral window alows flux from the model
to escape from deeper, hotter layers of the atmosphere
than in the real planet. }
\label{fig-2}
\end{figure}
neighboring wavelengths which arise deeper
in the atmosphere where the model more closely
follows the actual temperature profile agree much
better with the observations.  

Mid-infrared spectra of extrasolar giant planets and brown dwarfs will thus provide
substantial information about the temperature profile above the tropopause and 
consequently information on stratospheric heating mechanisms.

\section{Condensation}  
\subsection{Solar Jovian Atmospheres}

The visual appearance of the solar jovian planets is predominantly
controlled by their tropospheric cloud decks and stratospheric hazes.
Clouds condense when the partial pressure of
a minor constituent exceeds the saturation vapor pressure at a given location
in the atmosphere.  Figure 1 includes the temperature-pressure
profile for Jupiter as well as  the vapor
pressure curve for water. Clouds are assumed to form where the two
lines intersect.  It is on the
basis of this sort of calculation that the composition of the clouds in jovian
atmospheres are inferred.

In Jupiter's atmosphere the uppermost cloud deck consists of ammonia 
ice particles.  Beneath this cloud is an optically thick layer of condensed
{$\rm NH_4SH$} which in turn likely overlies a water cloud.  The global extent
of the {$\rm NH_4SH$} cloud layer is revealed by center-to-limb
brightness variations across the disk (West 1979; Chanover 1997) which
are inconsistent with a single cloud layer.  However, on a local scale the
clouds are clearly patchy.  The nephelometer on the Galileo atmosphere probe, which penetrated
the atmosphere in a relatively
cloud free region, did not detect the thick clouds present over
most of Jupiter's disk (Ragent et al. 1996).

Rock forming elements, such as Fe, Si, Ti presumably are condensed as grains
far below the visible atmosphere and thus are generally not present
in Jupiter's visible atmosphere (Fegley \& Lodders 1994).

The condensation sequence in Saturn's atmosphere is similar to that
at Jupiter.  The atmospheres of Uranus and Neptune, however, are
substantially colder and their upper cloud layer is instead
compsed of condensed methane.
Below this cloud layer is likely an $\rm H_2S$ cloud deck, followed at
depth by a sequence similar to that seen at Jupiter and Saturn.

Clouds substantially affect both the appearance and spectra of the
jovian planets.
First, condensation removes species from the
atmosphere above the clouds.  It is generally expected that the abundance
of condensible species follows a saturation vapor pressure curve
above its own cloud tops.
While this is approximately true in general,
the actual abundance of condensible species above clouds in the
solar system is a complex issue (e.g. Lunine and Hunten 1989).
Secondly, clouds form reflecting layers which scatter incident
solar radiation.  An absorbing, Rayleigh-scattering atmosphere with
no clouds would look substantially different.  Photons streaming
downwards at red and longer wavelengths would generally be absorbed
before scattering, suppressing the flux in these
regions of the spectrum.
Thirdly, clouds  also may affect the emitted radiation.
At Jupiter, most of the thermal emission arises at or above the
ammonia cloud tops.  However emission in the $5 \,\rm \mu m$ spectral
window arises from deeper, hotter layers of the atmospheres.  A
comparison of Hubble Space Telescope visual and IRTF $5\, \rm \mu m$
images of the planet (Chanover 1997) reveals that there is a one-to-one
correspondence between holes in the ammonia cloud deck and regions
of five-micron emission.

Condensates of low-abundance species produced by photochemistry
are also found in all the jovian atmospheres.
In Jupiter's stratosphere methane and ammonia are photolyzed by
solar UV radiation.  A rich photochemistry ensues that results
in compounds such as ethylene ($\rm C_2H_4$) and acetylene ($\rm C_2H_2$) which
condense in the stratosphere, producing an optically thin haze
which lies above the ammonia cloud (West 1979).  Hazes  are particularly apparent
in spectral regions of strong methane absorption in the
near-infrared (e.g. the $2.3 \,\rm \mu m$ band).  In these methane
bands a portion of the  incident solar radiation that 
would  otherwise be almost entirely
absorbed  instead
 scatters off the hazes, thereby increasing  reflected flux.  Thus a small 
 column abundance of hazes  can substantially affect a  planet's spectra
 as well as the stratospheric temperature.

\subsection{Extrasolar Jovian Atmospheres}
Condensation will also play an important role in controlling the
properties of the EGPs.  Figure 1 shows
model temperature-pressure profiles for a selection of the
extrasolar planets.  Of the selected atmospheres, water clouds would
be expected to condense in the atmospheres of 47 UMa b and 55 Cnc A c, but not
in the atmospheres of 70 Vir b and HD 114762 b.  

In atmospheres that are too warm for water condensation, there
are few other condensates until about 2000 K when refractory oxides
such as CaTiO$_3$ and $\rm Mg_2SiO_4$ condense.  As shown
in Figure 3, such condensates will form well below the observable
\begin{figure}
\plotone{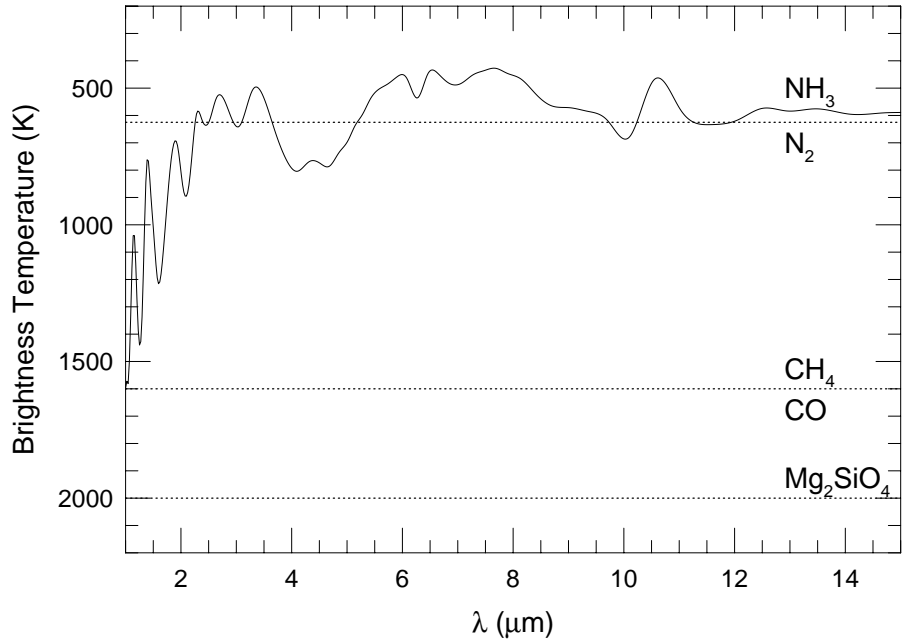}
\caption{Brightness temperature spectrum for baseline
Gl 229 B model of Marley et al. (1996).  Temperature is
plotted increasing downwards to suggest depth at which
flux originates in the atmosphere.  Physical temperature
at which NH$_3$ - N$2$ and CH$_4$ - CO are in thermochemical
equilibrium are shown by horizontal lines.  Also shown is
temperature at which enstatite condenses.  The 
radiative-convective boundary in this model is quite deep, below 1700 K.  }
\label{fig-3}
\end{figure}
atmosphere of Gl 229 B.  Thus atmospheres in a temperature
range from somewhat warmer than Jupiter to somewhat warmer than
Gliese 229 B (or $400 < T_{\rm eff} < 1200\,\rm  K$  (Guillot et al. 1996))
will lack abundant condensates. Trace compounds
such as ZnS and $\rm Na_2S$ may condense in these atmospheres, 
forming low optical depth
hazes.  Guillot et al. (1996) explore the condensation of trace species
further.

Approximate reflected visible spectra for several extrasolar jovian planets
and Jupiter are shown in Figure 4.  There is no cloud assumed
\begin{figure}
\plotone{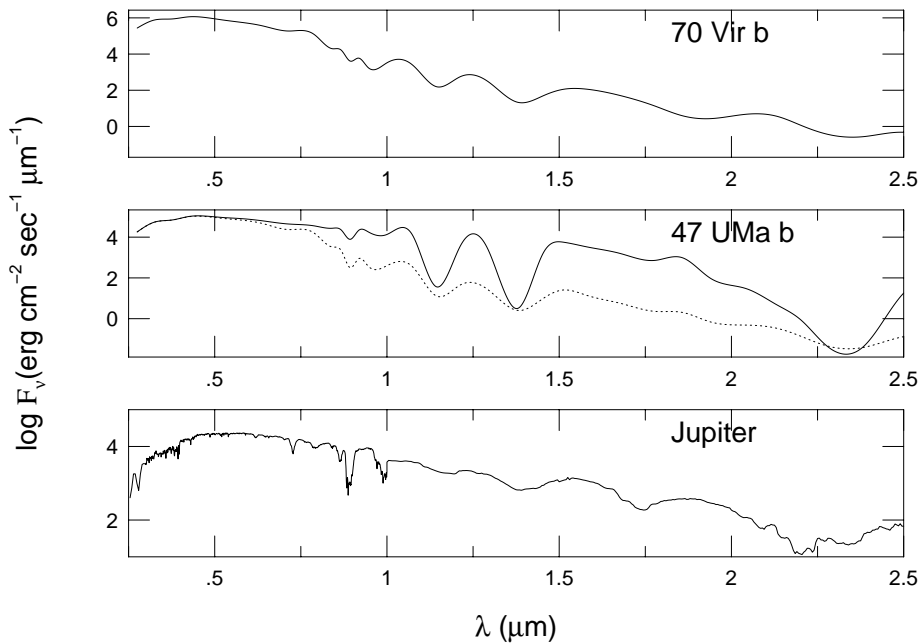}
\caption{Model reflected visible and near-infrared spectra for 70 Vir b, 47 UMa b, 
and observed Jupiter spectrum.  For 47 UMa b two models
are shown, one with (solid) and one without (dotted) clouds.  
The EGP model spectra are preliminary, but demonstrate
the substantial sensitivity to clouds.  Note the remarkable
fall off of reflected flux in the near-infrared in the cloud-free models.
}
\label{fig-4}
\end{figure}
for the atmosphere of 70 Vir b.  For 47 UMa b, two models are shown, one with
and one without clouds.   The water cloud is simply modeled as
a conservatively-scattering grey cloud of optical depth 10,
which is typical for terrestrial clouds.  The presence or
absence of clouds substantially alters the near-infrared reflected
spectrum for 47 UMa b.  In the absence of clouds, stellar photons from the
primary propagate downwards and are typically absorbed before
they Rayleigh scatter.  Thus the reflected flux beyond 1 micron
is quite small.  With a cloud layer many more of the photons scatter
before they are absorbed and the planet is much brighter in this
spectral region.

Clearly the near-infrared reflected flux of the extrasolar jovian
planets can provide a first-order measure of the presence of
atmospheric condensates.  This in turn provides a measure of the
atmospheric temperature.

Such a test however, will be complicated by the emitted thermal
radiation.  For objects warmer than about  400 K, the emitted flux
in some bands will exceed the reflected flux in the 1 to 3 micron region
for planets at several AU from
typical primaries.  If spectra of such objects are obtained,
detailed  modeling of the reflected and thermal components will
be required.

\subsection{Brown Dwarfs}

In the atmospheres of the cooler brown dwarfs condensation
is important more from a chemical perspective than
a radiative perspective.

At temperatures near 2000 K in the atmosphere of Gl 229 B, iron and
various refractory silicates condense (Fegley and Lodders 1996).  
Thus in a rising parcel of gas,
these materials will form grains and perhaps a discrete layer.  The exact
behavior  will depend on the grain size, grain growth rate, and convective velocities.
The primary consequence of grain formation is that refractory elements, such
as Ti, V, and Fe are removed from the gaseous atmosphere.  For
example Ti condenses as $\rm CaTiO_3$, or perovskite.  Vanadium
dissolves into perovskite and is likewise removed (Fegley and Lodders
1994).  Indeed,
absorption lines of refractory diatomic molecules, such as TiO are VO are
not observed in the atmosphere of Gliese 229 B (e.g. Allard et al. 1996).  

The grains themselves, however, are expected to be hidden from view
in the atmosphere of Gl 229 B.  As shown in Figure
3, the grain formation region (represented in the figure by 
$\rm Mg_2SiO_4$ lies well below the observable atmosphere.
Unless grains are convectively transported to the visible portion
of the atmosphere, they would not be expected to play a major role
in controlling the spectrum.
Since most of the acceptable Gl 229 B models of Marley et al. (1996) 
are radiative at some point between the region of
grain condensation and the photosphere, upward transport
of grains is unlikely.  As with the EGPs
discussed above, however, the condensation of minor species
may produce detectable hazes in the atmosphere.

\section{Non-equilibrium chemistry}

To an excellent first approximation, the atmospheres of the
solar jovian planets are in thermochemical equilibrium.  Thus
C, O, and N are predominantly found in the forms of CH$_4$,
H$_2$O, and NH$_3$, as would be expected at the temperatures
and pressures of the observable atmospheres.  
Fegley \& Lodders report on the abundances of
species expected at chemical equilibrium
in the atmospheres of Jupiter and Saturn (1994) and Gl 229 B (1996).

However the timescales
of convective motion in these planetary atmospheres are sufficiently
short compared to chemical equilibration timescales that some
non-thermochemical equilibrium species are present in the
observable atmospheres of these planets.  Species which
equilibrate at higher temperatures and pressures deeper
into the planet are dredged up to the visible atmosphere
by convection.

An example of this process is CO at Jupiter.  The carbon monoxide
mixing ratio measured in Jupiter's atmosphere, $1.6\times10^{-9}$ 
(Noll et al. 1988), is more than ten orders of magnitude (Fegley \& Lodders
1994) larger than the chemical equilibrium abundance.  This discrepancy
is explained by the convective transport of CO from depths where
the temperature is about 1100 K and the observed abundance of CO is in
equilibrium with water and methane.

In elementary mixing length theory (e.g. Clayton 1968) the 
convective velocity is proportional to the cube root of
the convective flux.  Thus in Jupiter-mass and larger planets
of comparable or lesser age, convective velocities will be
comparable to those in Jupiter or greater.  Convective
transport of non-equilibrium species will likely also
be an important process in these atmospheres.

Whether or not convective transport is an important process
in the atmosphere of Gl 229 B depends on the mass and gravity.
For some models, such as the baseline model of Marley et al.
illustrated in Figure 3 ($T_{\rm eff}= 960\,\rm  K$ and $g=1000\,\rm
m\,sec^{-2}$), the
radiative/convective boundary is quite deep, below 1700 K
and the point at which CO is in chemical equilibrium with $\rm CH_4$.
Since the atmosphere is radiative above this point, there
would be no upward transport.  From Figure 3
it is apparent that the CO 1-0 vibration-rotation band
near $4.7\,\rm \mu m$ 
(where CO has been detected in the jovian planets (Noll et al. 1988))
probes to temperatures no deeper
than about 800 K and thus CO would be expected to virtually undetectable.

However in other somewhat cooler and less massive models a
detached convection zone is predicted to appear near 1 bar.
Thus convection could mix air from around 1100 K to the observable
atmosphere around 800 K, bringing with it detectable, non-equilibrium,
amounts of CO.

Noll and Marley (1997) consider the detectability of
non-equilibrium species in Gl 229 B.  CO is the most
easily detected non-equilibrium species.  $\rm PH_3$
is also potentially detectable.  If CO is detected, the implied
presence of an upper, detached convection zone would help constrain
the atmosphere model.

\section{Deep Radiative Zones}

Prior to work by Guillot et al. (1994), it had long been assumed
that the deep atmospheres of the solar jovian planets were fully convective at
all depths below the radiative-convective boundary, slightly
below the photosphere.  The large
pressure-induced  opacity   of $\rm H_2$ as well as the molecular opacities
of water, methane, and ammonia, along with free-free and bound-free absorption
at depth, would presumably require steep temperature gradients to transport
heat by radiation.  Since both the conductivity and viscosity of dense hydrogen-helium
fluids are low, it was assumed that the adiabatic temperature gradient
would always be less steep than the radiative-equilibrium lapse rate and
the atmospheres would convect.

However Guillot et al. (1994) demonstrated that the  near-infrared windows in the
important opacity sources from 1 to $2.4\,\rm \mu m$ would create a region 
from about 1,000 K to about
3,000 K through which a substantial amount of radiation could be transported.
The boundaries of this region are controlled by the overlap
of the Planck function with the windows in methane, water, and pressure-induced
hydrogen opacity.   At lower temperatures the
Planck function does not substantially overlap the window regions and
energy transport is again by convection.  At higher temperatures the electron
number density grows large enough that free-free and bound-free
absorption closes the windows.

The models of Marley et al. (1996)  predict that this same process is
 at work in the atmospheres of brown dwarfs.
However since these objects are warmer, the radiative region arises
at much lower pressures than on Jupiter.   Thus in models where
it is present, the deep radiative zone is found much closer to
the observable atmosphere than on Jupiter.  Thus its effects
may be more important.

The radiative window is important because it disconnects the upper convection
zone from the deep interior.  Planetary evolution models require as a boundary
condition a temperature and pressure within the deep convection zone.  Thus models
must account for the presence of the radiative window.  Furthermore if convection indeed
delivers non-equilibrium  constituents to the observable atmosphere in the presence of
the two zones, the upper zone will dredge only to the bottom of that zone.

\section{Summary}

When theorists first approach the a new class of objects, 
like the extrasolar giant planets and the cool brown dwarf Gl 229 B,
they are tempted to treat it as a ``homogeneous, spherical object''.
However the exploration of the solar system has clearly taught us
that planets are complex, unexpected objects.  Processes which
at first seem second order in importance may control first order
observable characteristics.  For example Jupiter might be
expected to have thick, uniform
water and ammonia cloud decks, which could prevent any flux from
emerging from hot, deep seated regions of the atmosphere. 
Yet Jupiter's substantial flux in the five-micron window
in molecular opacities emerges through holes in the clouds.

We should not expect the atmospheres of the EGPs and brown
dwarfs to be any less complex.  The selection of processes
discussed here will likely play important roles in at least
some of these new atmospheres.  The exploration of the solar system
has prepared us to learn from 
these new worlds.  Doubtless these worlds will teach us new
lessons about unexpected processes that shape their atmospheres.

% Finally, we have a little acknowledgements section.

\acknowledgments

This work was supported by NASA grant NAG2-6007 and NSF grant
AST-9624878.

% That's the end of the main body of the paper.  Now we will have some
% back matter.

% Now comes the reference list.  Since we typed out the citations ourselves,
% the reference list is enclosed in a "references" environment.  Each
% new reference begins with a \reference command which sets up the proper
% indentation.  Typography that may be required in the reference list by
% the editorial staff must be included by the author.
%
% Observe the "standard" order for bibliographic material: author name(s),
% publication year, journal name, volume, and page number for articles.
% Some journal names are available as macros; see the WGAS markup
% instructions for a listing of which ones have been "macro-ized".
% Note the use of curly braces to delimit the font changes: it is essential
% that this be done to limit the scope of the font declaration.
%
% There is no need to engage in any other typographic manipulation.

% That's all, folks.
%
% The technique of segregating major semantic components of the document
% within "environments" is a very good one, but you as an author have to
% come up with a way of making sure each \begin{whatzit} has a corresponding
% \end{whatzit}.  If you miss one, LaTeX will probably complain a great
% deal during the composition of the document.  Occasionally, you get away
% with it right up to the \end{document}, in which case, you will see
% "\begin{whatzit} ended by \end{document}".


\begin{references}
\reference Allard, F., Hauschildt, P., Baraffe, I., and Chabrier, G., 1996,
Ap. J. 465, L123.
\reference Appleby, J. 1986. Icarus, 65, 383.
\reference Butler, P., Marcy, G., Williams, E., Hauser, H., and Shirts, P.
1997, \apj, 474, L115.
\reference Chamberlain, J. and Hunten, D. 1987. Theory of Planetary Atmospheres
(San Diego:Academic Press).
\reference Chanover, N. 1997, Temporal Variations in the Vertical
Structure of Jupiter's Atmosphere, Ph.D. thesis, NMSU.
\reference Clayton, D. 1968, Principles of Stellar Evolution and Nucleosynthesis
(New York:McGraw Hill).
\reference Fegley, B. and Lodders, K. 1994. Icarus, 110, 117.
\reference Fegley, B. and Lodders, K. 1996. \apj 472, L37.
\reference Goody, R., West, R., Chen, L., and Crisp, D. 1989. J. Quant.
Spectr. Rad. Transfer, 42, 539.
\reference Grossman, A. and Grant, K. 1994. Eighth Conf. on Atmos. Radiation,
 American Met. Soc., 97.
\reference Guillot, T., Gautier, D., Chabrier, G., and Mosser, B. 1994. Icarus 112, 337.
\reference Guillot, T., Saumon, D., Burrows, A., Hubbard, W., Lunine, J.,
Marley, M., and Freedman, R. 1996,
in Astronomical and Biochemical Origins and the Search for Life 
in the Universe, eds. C.B. Cosmovici, S. Bowyer and D. Werthimer, 
Editrice Compositori, 343.
\reference Lacis, A. and Oinas, V. 1991. J. Geophys. Res., 96, 9027.
\reference Lindal, G. et al.  1981, JGR, 86, 8721.
\reference Lunine, J. and Hunten, D. 1989. Planet. Space Sci., 37, 151.
\reference Marley, M., Saumon, D., Guillot, T., Freedman, R., Hubbard, W.,
Buttows, A., and Lunine, J. 1996, Science, 272, 1919.
\reference Marley, M., McKay, C., and Pollack, J. 1997, Icarus, in press.
\reference Noll, K., Knacke, R., Geballe, T., and Tokunaga, A. 1988, Ap.J. 324, 1210.
\reference Noll, K. and Marley, M. 1997. in Planets Beyond the Solar System,
ed. D. Soderblom (San Francisco; ASP), in press.
\reference Ragent, B., Colburn, D., Avrin, D., Rages, K. 1996. Science 272, 854.
\reference Rages, K., Pollack, J., Tomasko, M., and Doose, L. 1991, 
Icarus, 89, 359.
\reference Noll., K. and Marley, M. 1997. Planets Beyond the Solar System 
and the Next Generation of Space Missions, ed. D. Soderblom, ASP
Conf. Series, in press.
\reference West, R. A. 1979, Icarus, 38, 34.
\end{references}
\end{document}